\documentclass[aps,pre,superscriptaddress,twocolumn]{revtex4}
\usepackage[dvips]{graphicx}
\usepackage{amsmath}
\usepackage{amssymb}
\usepackage{natbib}
\usepackage{psfrag}
\usepackage{color}
\usepackage{epsfig}
\usepackage{subfigure}

\begin{document}

\title{Stochastic timing in gene expression for simple regulatory strategies}

\author{Alma Dal Co}

\affiliation{Department of Physics and INFN, Universit\`a degli Studi di Torino, via P.Giuria 1, 10125 Turin, Italy.}

\affiliation{Current address: ETH Z\"urich, Department of Environmental Systems Sciences, Universit\"atstrasse 16, 8092 Z\"urich, Switzerland, and Eawag, Department of Environmental Microbiology, \"Uberlandstrasse 133, 8600 D\"ubendorf, Switzerland.}

\author{Marco Cosentino Lagomarsino} 

\affiliation{Sorbonne Universit\'es, Universit\'e Pierre et Marie Curie, Institut de Biologie Paris Seine, Place Jussieu 4, Paris, France.}

\affiliation{UMR 7238 CNRS - Universit\'e Pierre et Marie Curie, Computational and Quantitative Biology, Rue de l'Ecole de Médecine 15, 75006 Paris, France.}

\affiliation{IFOM, FIRC Institute of Molecular Oncology, Via Adamello 16, 20139 Milan, Italy.}

\author{Michele Caselle} 
\affiliation{Department of Physics and INFN, Universit\`a degli Studi di Torino, via P.Giuria 1, 10125 Turin, Italy.}

\author{Matteo Osella} 
\email[correspondence to: ]{mosella@to.infn.it}
\affiliation{Department of Physics and INFN, Universit\`a degli Studi di Torino, via P.Giuria 1, 10125 Turin, Italy.}


\date{\today}

\begin{abstract}
Timing is essential for many cellular processes, 
from  cellular responses to external stimuli to the cell cycle and 
circadian clocks. Many of these processes are based on gene
expression. For example, an activated gene may be required to reach in a precise time a   
threshold level of expression that triggers  
a specific downstream process. 
However, gene expression is subject to stochastic fluctuations, 
naturally inducing an uncertainty
in this threshold-crossing time with potential consequences on biological functions and
phenotypes. Here, we consider such ``timing fluctuations'', 
and we ask how they can be controlled. Our analytical
estimates and simulations show that, for an induced gene,  
 timing variability is minimal if the threshold 
level of expression is approximately half of the steady-state
level. Timing fluctuations can be reduced by increasing the
transcription rate, while they are insensitive to the translation
rate. In presence of self-regulatory strategies, we show that
self-repression reduces timing noise for threshold levels that
have to be reached quickly, while self-activation is optimal at
long times. These results lay a framework for understanding
stochasticity of endogenous systems such as the cell cycle,
as well as for the design of synthetic trigger circuits.

\end{abstract}

\maketitle

\section{Introduction}

Several cellular processes rely on a precise temporal
organisation~\cite{Pedraza2007,Yurkovsky2013}.  Prominent examples are
the controls of the cell cycle and of circadian clocks, where the
timing precision can be crucial for the correct cellular
physiology~\cite{Pedraza2007,Bean2006}.  Similarly, the complex
patterns of sequentially ordered biochemical events that are often
observed in development and cell-fate decision presumably require a
tight control of expression
timing~\cite{Eldar2010,Yosef2011,Kuchina2011}.  Typically, internal
signals and environmental cues induce the expression of one or several
regulators, which in turn can trigger the appropriate cellular
response when their concentration reach a certain threshold
level~\cite{Pedraza2007,Singh2014}.  

However, a gene may reach a target level of expression with
substantial cell-to-cell variability, even in a genetically identical
population of cells exposed to the same stimulus. This variability is
a necessary consequence of the intrinsically stochastic nature of gene
expression~\cite{Paulsson2004,Raj2008}.
For genes whose expression has to reach a trigger threshold level,
noise in gene expression leads to variability in the time required to
reach the target level.  This raises the question of what is the
extent of this variability and which regulatory strategies can control
such fluctuations.  Most studies have focused on fluctuations in
molecule numbers at equilibrium, while comparatively very few studies
have addressed the problem of timing fluctuations
theoretically~\cite{Murugan2011,Singh2014,Yurkovsky2013} or
experimentally~\cite{Amir2007,Pedraza2007,Nachman2007}.

Here, we develop analytical estimates and simulations to study the
fluctuations in the time necessary to reach a
target expression level after gene induction, and we investigate the effect of simple 
regulatory strategies on these fluctuations.  
We first consider the case of an unregulated gene whose expression is switched on,    
and we ask what are the 
relevent parameters defining the crossing-time fluctuations and how these fluctuations 
can eventually be reduced by the cell. 
Second, we investigate the role of simple regulatory
strategies in controlling the expression timing fluctuations, focusing
on the two circuits of positive and negative transcriptional
self-regulation.

Understanding expression timing variability is
key to approach basic biological mechanisms at the single-cell
level. Isolating the possible regulatory strategies able to
control this variability can be useful to decipher the design principles
behind regulatory networks associated to cellular timing. 

\enlargethispage{-65.1pt}

 With todays experimental techniques based on fluorescence time-lapse microscopy~\cite{Young2012}, 
 potentially coupled with microfluidic devices  to keep cells in a controlled environment for many 
 generations~\cite{Bennett2009,Long2014}, data on gene expression timing and its fluctuations will be more and 
 more accessible in the next years. Therefore, a parallel theoretical understanding of timing fluctuations 
 is necessary to interpret this upcoming data, to design focused experiments, and eventually to engineer 
 synthetic circuits with specific timing properties.

\section{MATERIALS AND METHODS}

\subsection{Background on the ``standard model'' for gene expression}
\label{sec:model}

  \begin{figure}
  \begin{center}
H \includegraphics[width=0.4\textwidth]{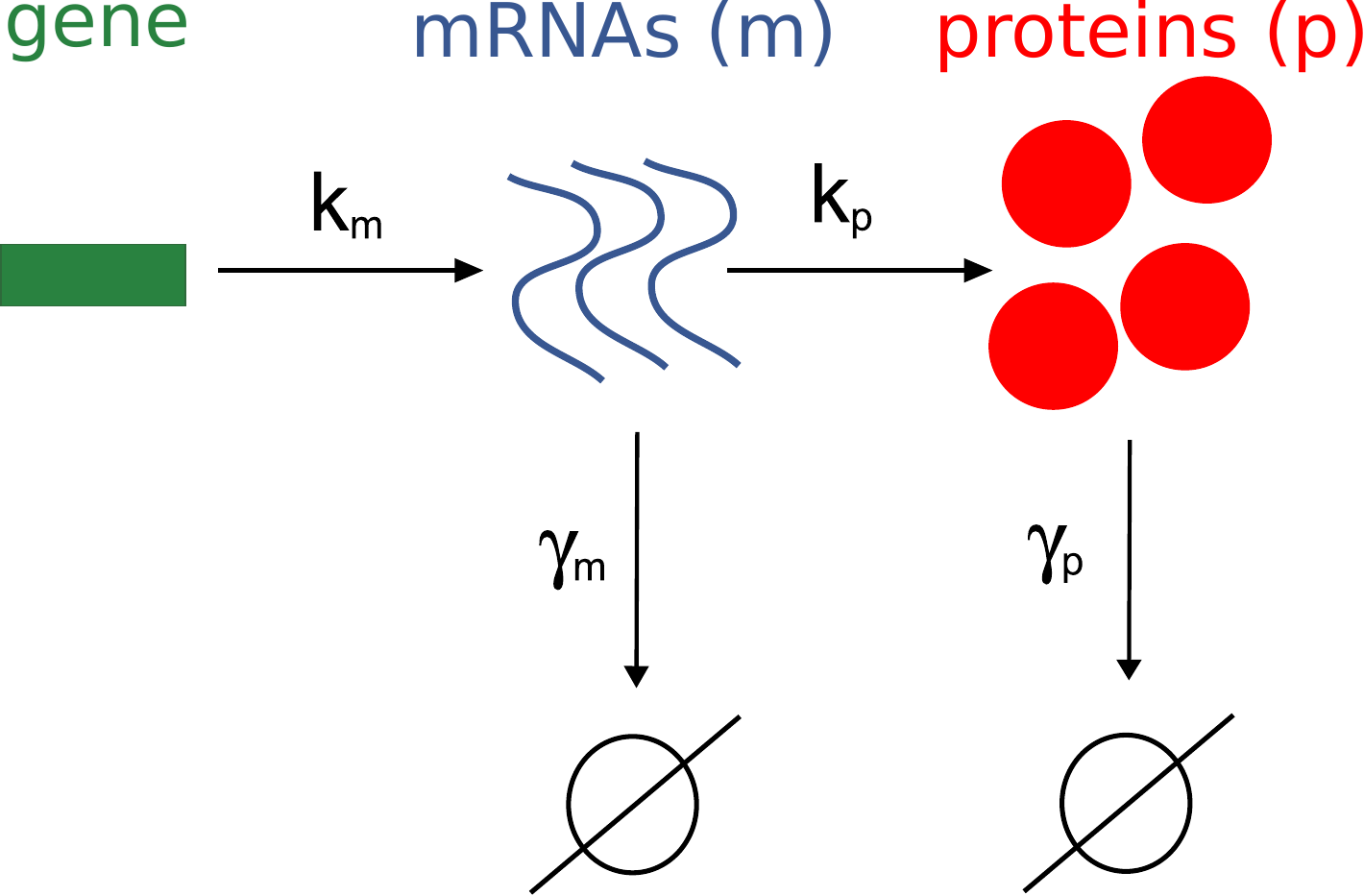}
 \end{center}
\caption{\textbf{Model of gene expression.}  A basic description of gene
   expression includes transcription, translation and molecule
   degradation.  The model described by Eqs.~\ref{eq:mean_field}
   and their stochastic version is based on this scheme.  }
  \label{fig1}
  \end{figure}

  We employed a standard model of stochastic
  gene expression~\cite{Paulsson2005} (Fig.~\ref{fig1}) taking into account messenger RNA (mRNA)
  and protein production and
  degradation
  as first-order chemical reactions (with rates
  $k_m, k_p$ for productions and $\gamma_m,\gamma_p$ for
  degradations)~\cite{Paulsson2005,Kaern2005,BarEven2006,Shahrezaei2008}.  The rate equations describing
  the average mRNA and protein dynamics are
\begin{eqnarray}
\frac{d {m(t)}}{{d t}} & =& k_{m} - \gamma_{m} m(t) \nonumber\\
\frac{d {p(t)}}{{d t}} &=& k_{p} m(t) - \gamma_{p} p(t) .
\label{eq:mean_field}
\end{eqnarray}

Since  the gene expression process in Fig.~\ref{fig1} does not entail any transcriptional or post-transcriptional regulation we refer to it  as ``constitutive'' expression in the following.   
However, we are interested in the activation dynamics  to evaluate the time (and its fluctuations) necessary to reach a certain level of expression (Fig.~\ref{fig2}). 
Thus, the case of a step induction of transcription will be considered, following e.g. ref.~\cite{Rosenfeld2002}. 
The  kinetics after a step induction is modeled by the solution of Eqs.~\ref{eq:mean_field}  with initial conditions $m(0)=0$ and $p(0)=0$

\begin{eqnarray}
p(t) &=& p_{ss} \left( \frac{\gamma_p (1-e^{- \gamma_m t})-\gamma_m (1-e^{- \gamma_p t})} {\gamma_p - \gamma_m}\right)\nonumber\\
&\simeq& p_{ss} (1-e^{-\gamma_p t}),
\label{sol:mean_field}
\end{eqnarray}

where  $p_{ss}= k_p k_m / \gamma_m \gamma_p$ is the protein steady-state value, and the approximation holds 
for a protein halflife much longer than the mRNA halflife, i.e., $\eta= \gamma_p/\gamma_m \ll 1$. 
Indeed, especially in microorganism such as bacteria and yeast, proteins are typically stable, with a lifetime longer
than the cell cycle, while mRNAs have a lifetime of just  few minutes~\cite{Taniguchi2010,Shahrezaei2008}, justifying the assumption $ \eta \ll 1$~\cite{Thattai2001,Shahrezaei2008}.  
Moreover, the loss of highly stable proteins, captured by the rate $\gamma_p$,  is  
mainly due to dilution through growth and cell division, so that an
effective degradation rate $\gamma_p = \mu~ln2$ (where the growth rate $\mu$ is the
 the inverse of the cell doubling time)  can be safely assumed in most cases~\cite{Osella2013}. 

The threshold level $\tilde{p}$ of protein expression to be crossed was defined in units of the steady-state value of expression, 
with the dimensionless parameter $\alpha=\tilde{p}/p_{ss}$ (Fig.~\ref{fig2}) .
Given the  threshold $\alpha$, the corresponding average crossing time can be numerically calculated from Eq. ~\ref{sol:mean_field}, 
while it takes the simple form  $t\simeq - log(1-\alpha)/\gamma_p $  for $\eta \ll\ 1$.

The master equation controlling the time evolution of  
the  probability of having $m$ mRNAs and $p$ proteins at time $t$ can be solved analytically for constitutive  expression~\cite{Shahrezaei2008}.  
Since the mRNA  dynamics is described by a birth-death process, 
the distribution of mRNA numbers is Poisson.  By contrast, protein abundance  follows a broader
distribution, because of the amplification of mRNA fluctuations by 
protein translation bursts. The model predicts at steady state 
a negative binomial distribution~\cite{Shahrezaei2008}, and a gamma
distribution in the limit of $p$ continuous~\cite{Friedman2006}.
Fluctuations in protein number can be  measured by the  Coefficient of Variation (CV)  
which is the ratio between the standard deviation and 
the average number of proteins $CV_p (t)= \sigma_p(t)/ \langle p(t) \rangle$. 
The time evolution of this noise measure  has a particularly compact form in the regime of $\eta\ll1$~\cite{Shahrezaei2008}:

\begin{equation}
 CV_p(t)^{2} = \frac{1}{\langle p (t)\rangle} (1+b +b e^{-\gamma_p t}), 
 \label{cvp}
\end{equation}

where $b= k_p/\gamma_m$ is the burst size, i.e.,  the average number of proteins produced during a mRNA lifetime, while the dynamics of 
$\langle p(t) \rangle$ is described by the deterministic equation (Eq.~\ref{sol:mean_field}).
The burst size  represents the amplification 
factor of noise with respect to the Poisson noise of the mRNA. Indeed, the noise expression at steady state   
$CV_p^{2} = \frac{1}{\langle p\rangle_{ss}} (1+b)$ explicitily shows  a noise term proportional to $b$ in addition to the Poisson scaling $\sim1/\langle p\rangle$.
A ``burst frequency'' parameter $a= k_m /\gamma_p$ can be defined  
so that the average protein level at steady state is expressed as the product $\langle p\rangle_{ss}= a~b$.

\subsection*{Estimate of biologically relevant parameter values}

The biologically relevant range of parameters can be
extrapolated from large-scale measurements of gene expression at the
single cell level.  For \textit{E. coli} in particular, proteins and
mRNAs have been measured with single-molecule sensitivity for $\sim
10^3$ genes~\cite{Taniguchi2010}.  In this dataset, the average mRNA
lifetime is $~5$ minutes, which corresponds to a degradation rate of
$\gamma_m = 0.2$.  Protein lifetime is often longer then the duration
of the cell cycle, which may span from $20$ minutes to several hours in
fast-growing bacteria like~\textit{E. coli}.  Thus, protein dilution
in fast-growth conditions defines the maximum value $\gamma_p\simeq
0.03$ min$^{-1}$ for effective protein degradation.  The higher
stability of proteins with respect to mRNAs (i.e., the
approximation $\eta\ll 1$) is generally valid in
yeast~\cite{Shahrezaei2008} as well as in mammalian
cells~\cite{Schwanhaeusser2011}, although in higher eukaryotes the
many layers of regulation of molecule stability can give rise to a
more complex scenarios.
 
The average number of proteins per cell in~\textit{E. coli} ranges
from less than a unit to thousands~\cite{Taniguchi2010}.  The
corresponding burst size and frequency have been estimated from
fitting the distributions of protein numbers for different genes with
a Gamma distribution~\cite{Taniguchi2010}, i.e., the model prediction
of the steady-state distribution of the stochastic process based on
the scheme in Fig.~\ref{fig1} in the continuous $p$ limit and for
$\eta\ll 1$ ~\cite{Friedman2006}.  
However, for highly expressed genes  
extrinsic noise is empirically the dominant 
noise source in~\textit{E. coli} ~\cite{Taniguchi2010}. 
In this case, the values of $b$ and $a$  obtained from fitting cannot be strictly interpreted as the burst size and frequency, 
since the underlying model does not include extrinsic fluctuations.
Nevertheless, the average protein number can be roughly approximated by the product $ab$, 
although  corrections due to extrinsic fluctuations can emerge~\cite{Shahrezaei2008b}, 
making our order-of-magnitude estimate of the biologically relevant parameter range still meaningful.
With this caveat in the interpretation of  $a$ and $b$,  empirically we find that these parameters span a broad
range. This makes the relevant parameters strongly gene dependent.  The burst size has a long-tail distribution, 
ranging from 1 to thousands of proteins translated in a mRNA
lifetime. The average value is around 21 proteins while the median is
3.  The burst frequency of active genes is close to 5 mRNAs
per cell cycle.

The rates of transcription $k_m$ and translation $k_p$ can be
explicitly calculated once the molecule lifetimes have been fixed.
For example, bursts of frequency $a=10$ and size $b = 5$ correspond to
transcription and translation rates of $k_m = 0.1$ min$^{-1}$ and 
$k_p= 1$ min$^{-1}$, if the mRNA degradation is set to $\gamma_m =
0.2$ min$^{-1}$ (i.e., the empirical average value) and the cell-cycle
time is around 70 minutes.  Additionally, all these parameters are influenced by cell physiology, and
in particular they are growth-rate dependent in bacteria~\cite{Klumpp2009}.

Although an extensive exploration of this large parameter space is not
feasible, we tested our model results with several parameter
sets inside this biologically relevant range, finding a qualitative
agreement.  The main figures are based on the example
of a gene expressing $2000$ proteins at
the steady state, a burst size and frequency chosen compatible with this steady 
state value, with the burst size laying in the range
$b\in[2,100]$ (setting the noise at the
protein level), mRNA lifetime around $5$ minutes,  effective protein 
lifetime set by the cell-cycle time.

\subsection*{Stochastic simulations}

Simulations were implemented by using Gillespie's first reaction
algorithm~\cite{Gillespie1976}. 
The stochastic reactions simulated
are those presented in Figure~\ref{fig1} for constitutive expression.
Reactions that depend on a regulator, as in the two self-regulatory
circuits, were allowed to have as rates the corresponding full
nonlinear functions (Eqs.~\ref{hill}).  Each data point in the figures
is the result of $10^4$ trials.

\section{RESULTS}

\subsection{Estimate of first-passage time fluctuations}

\begin{figure}
  \begin{center}
 \includegraphics[width=0.4\textwidth]{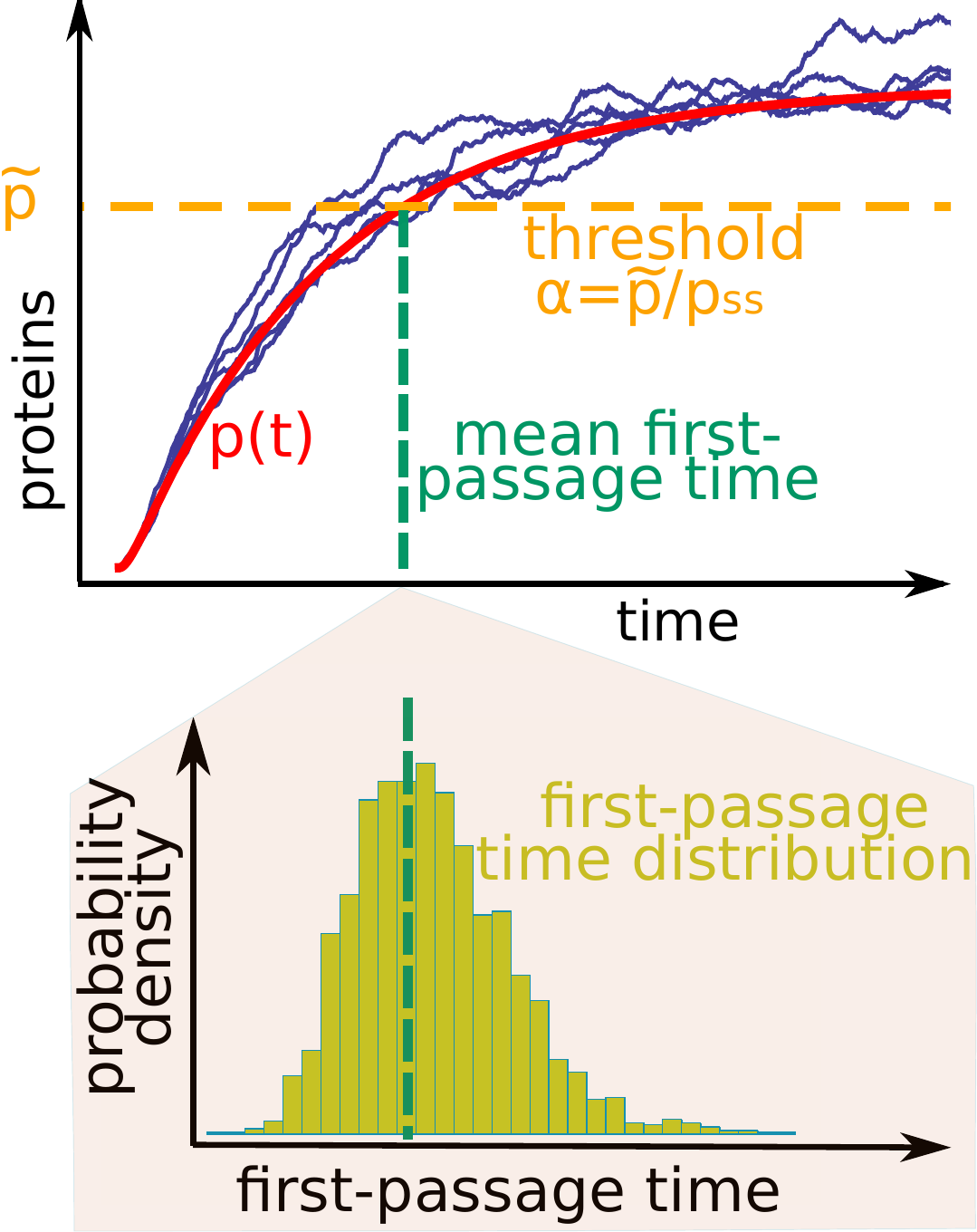}
 \end{center}
\caption{  \textbf{ Definition of the
     threshold-crossing problem.} When
   a  gene is switched on at time $t=0$ the average protein level 
   (red continuous line) approaches the steady state with the dynamics
   described by Eq.~\ref{sol:mean_field}.  Since gene expression is a
   stochastic process, individual trajectories fluctuate around the
   mean behavior, as illustrated by stochastic simulations (blue
   lines).  The distribution of times of crossing a fixed protein
   level $\tilde{p}$ (orange horizontal dashed line) is the
   first-passage time distribution (histogram) representing the
   variabilty in reaching a certain level of expression. The average
   first-passage time can be directly deduced from the deterministic
   mean dynamics (green vertical dashed line).}
  \label{fig2}
  \end{figure}

\begin{figure}[t]
\begin{center}
 \includegraphics[width=0.4\textwidth]{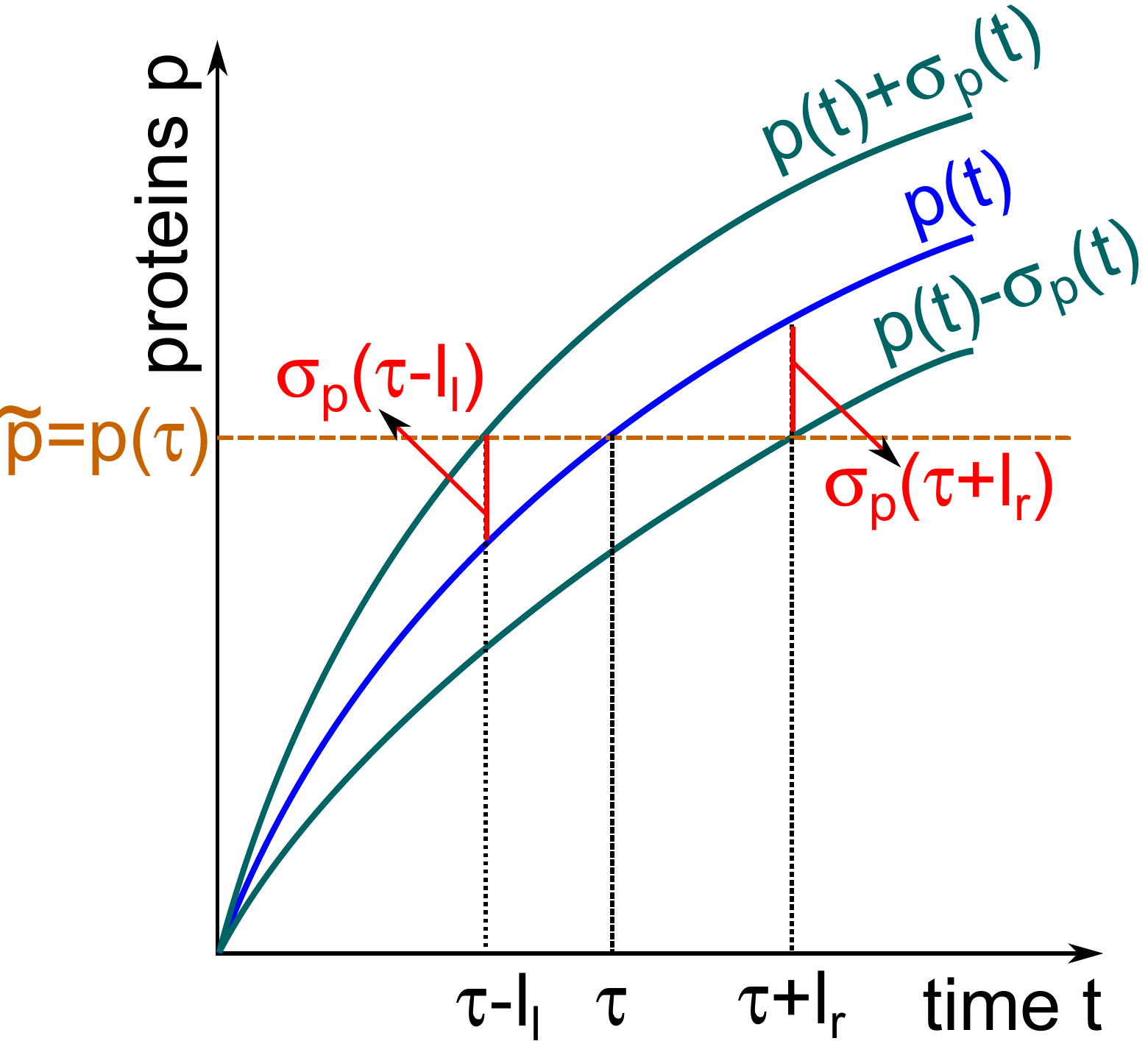}
 \end{center}
 \caption{  \textbf{Scheme of the geometric relations at the basis of
     the analytical estimate of timing fluctuations.}  The average
   protein level $p(t)$ rises after induction
   at time $t=0$. The dynamics depends on the specific
   regulations acting on the gene.  The time evolution of protein
   fluctuations is captured by the dynamics of the region at one
   standard deviation from the mean, defined by the curves $p(t)\pm
   \sigma_p(t)$.  The time necessary to these curves to cross a fixed
   threshold $\tilde{p}$ gives an estimate of the variability in the FPT.
  }
  \label{fig3}
\end{figure}

The threshold-crossing problem for gene expression is represented in Fig.~\ref{fig2}. 
After induction, the level of gene expression rises with a specific dynamics $p(t)$ and approaches the steady-state $p_{ss}$ for long times.   
We want to evaluate the time necessary to reach a target level of expression $\tilde{p}$, and  in particular its fluctuations. 
Mathematically, the problem of determining the time required for a
stochastic process to reach a certain value falls into the category of
 first-passage time (FPT) problems~\cite{Kampen2007}.
 Usually, FPT problems are difficult to treat analytically.  
Indeed, previous attempts of evaluating the FPT distribution in the context of
gene expression were based on numerical
approximations~\cite{Shahrezaei2008,Murugan2011}, or on simplified
processes, for example neglecting protein
degradation~\cite{Singh2014}.  
Here, we take a different approach. Our
goal is to provide a rough but simple analytical estimate of the FPT
noise that allows to identify and intuitively understand some of its
general properties.  To this aim, the geometrical arguments
illustrated in Fig.~\ref{fig3} can be used.  Fluctuations around the
dynamics of the average protein level $p(t)$, which for constitutive
expression is described by Eq.~\ref{sol:mean_field}, are quantified by
considering the region at a standard deviation distance from the mean
behavior, defined by the two curves $p(t) \pm \sigma_p(t)$ shown in
Fig.~\ref{fig3}.
A crossing threshold $\tilde{p}$ defines the dimensionless parameter
$\alpha=\tilde{p}/p_{ss}$ and the corresponding mean FPT $\tau$. The
two trajectories $p(t)\pm \sigma_p(t)$ cross the threshold $\tilde{p}$
at the time points $\tau-l_l$ and $\tau+l_r$ respectively.  The time
interval $[ \tau-l_l,\tau+l_r]$ can be considered as an estimate of
the variability in the FPT.  In particular, an approximation of the
standard deviation of the FPT is given by the average value
$\sigma_t(\tilde{p})\simeq\frac{l_r +l_l}{2}$, and this quantity can
be calculated explicitily as a function of the known parameters of the
process.
Figure~\ref{fig3} shows that two geometric relations hold: 

\begin{eqnarray}
 \sigma_p(\tau+l_r) ~&=&~ p(\tau+l_r) -p(\tau) \nonumber\\
  \sigma_p(\tau-l_l)~&=&~ p(\tau)-p(\tau-l_l) \ .
\label{general}
\end{eqnarray}

These expressions can be Taylor expanded around the mean FPT $\tau$

\begin{eqnarray}
 \sigma_p(\tau)  +  \frac{d}{dt} \sigma_p(t) \Big|_{\tau} l_r +...      ~&=&~   \frac{dp(t)}{dt}\Big|_{\tau} l_r +... \nonumber\\
  \sigma_p(\tau) -  \frac{d}{dt} \sigma_p(t)\Big|_{\tau}  l_l  +...    ~&=&~    \frac{dp(t)}{dt}\Big|_{\tau} l_l +...\ .
\label{approx}
\end{eqnarray}

Considering the first-order expansion and generalizing to all possible
threshold levels $p$ and their corresponding average FPT $t$, we
obtain a general relation between the variability in the FPT
$\sigma_t(p)$, the protein level variability at that time
$\sigma_p(t)$, and the average dynamics $p(t)$:

\begin{equation}
  \sigma_t(p)\simeq \frac{l_r +l_l}{2} \simeq  \sigma_p(t) \frac{dp(t)}{dt} \Bigl[\left(\frac{dp(t)}{dt}\right)^2 -\left(\frac{d\sigma_p(t)} {dt}\right)^2  \Bigr]^{-1}
\label{sigmat}
\end{equation}

If the variability in the protein level is approximately constant in
time, the above expression further simplifies to

\begin{equation}
\sigma_t(p) \simeq   \left(\frac{dp(t)}{dt}\right)^{-1} \sigma_p(t)
\label{sigmatzero}
\end{equation}
an equation reminiscent of the classic ``propagation of uncertainty'' in statistics.
Equations~\ref{sigmat} and \ref{sigmatzero} can be easily reformulated
in terms of the CV.  In particular, the lowest order approximation of
the CV of the FPT is

\begin{equation}
CV_t(p) \simeq \frac{ p(t)}{ t }  \left(\frac{d p(t)}{dt}\right)^{-1}  CV_p(t).  
\label{cvtzero}
\end{equation}

Note that $p(t)$ in this expression is the deterministic average
dynamics of the process, and $t$ is the average time if takes for
$p(t)$ to reach a generic fixed value $p$.  This relation between
noise in the protein level and noise in the timing of a threshold
crossing is quite general (although approximate), since it does not
require particular assumptions about the process in analysis.
Clearly, the caveat is that the time dependence of noise in the
protein level have to be known, which can be a severe limitation given
that an exact analytical solution is only known for constitutive
expression~\cite{Shahrezaei2008} (Eq.~\ref{cvp}), while it has to be
evaluated through numerical simulations even for simple regulatory
schemes.

\subsection*{Optimal out-of-equilibrium protein level for time measurements}

\begin{figure*}[t]
  \begin{center}
    \includegraphics[width=0.9\textwidth]{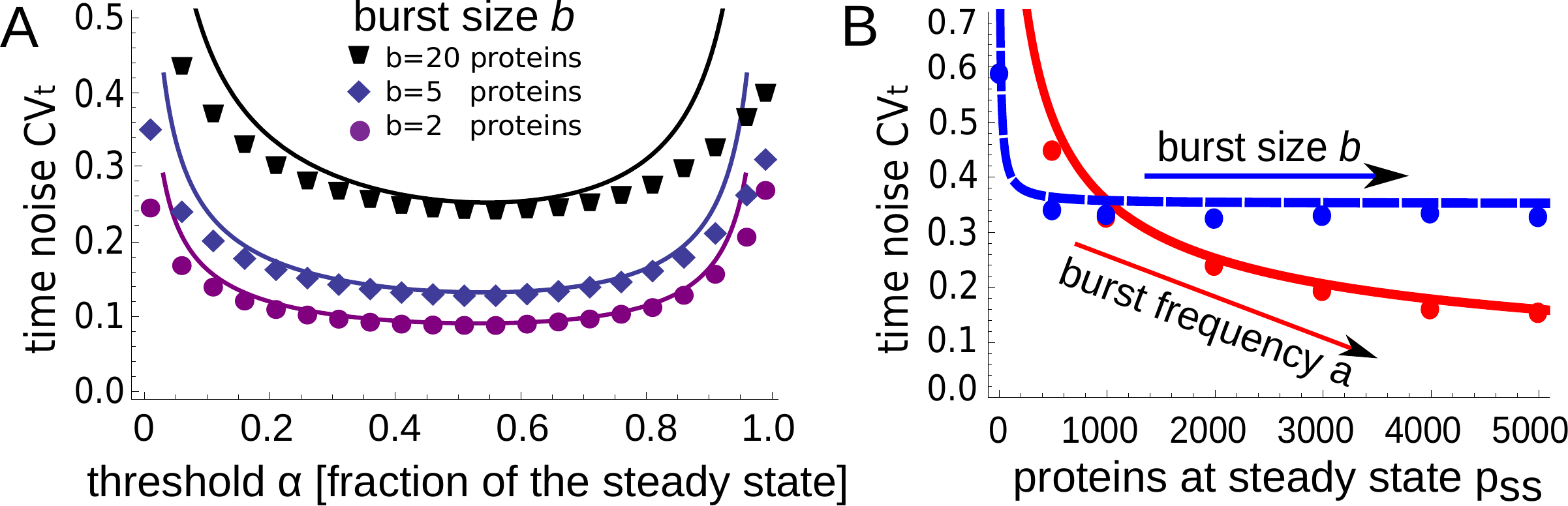}
  \end{center}
\caption{\textbf{Minimum of noise in the first-passage time at
     intermediate protein levels.}  A) The coefficient of variation
   $CV_t$ of the FPT is reported as a function of the target
   expression level $\alpha$ (in units of the steady-state value). The
   analytical estimate in Eq.~\ref{cvtzero} (continuous lines)
   predicts with good precision the results of exact Gillespie
   simulations (symbols). Both curves show a minimum timing noise
   value for an intermediate level of expression, around $\alpha\simeq
   0.55$.  We verified that the presence of a minimum is robust with
   respect to the parameter values, as expected from
   Eq.~\ref{cvtzero}.  The plot considers different values of the
   burst size $b$, the key determinant of protein number fluctuations
   (Eq.~\ref{cvp}), while keeping the steady state to a constant value
   ($ab=2000$ proteins).  The mRNA halflife is 5 minutes and the
   protein halflife around 8 hours, roughly corresponding to dilution
   for bacteria in slow-growth conditions (see the Materials and Methods section for the 
   biologically relevant range of the parameter values).
   B) Increasing the number of proteins produced (thus relaxing the
   constraint on the steady state value) lowers the noise in the
   timing fluctuations if expression is controlled at the
   transcriptional level, i.e., varying the burst frequency $a$.  On
   the other hand, increasing the translation rate, and thus the burst
   size $b$, does not affect the timing noise. Lines represent the analytical 
   predictions while dots are the results of Gillespie
   simulations.}
  \label{fig4}
\end{figure*}

This section addresses the role of the
positioning of the threshold protein level in determining the FPT
fluctuations.  We consider the example of a transcription factor (TF)
whose expression is turned on in response to an external stimulus.
Typically, the dependence of the target expression on the TF
concentration is sigmoidal~\cite{Alon2007,Bintu2005a}, triggering a response when the level of
expression approximately matches the dissociation constant of the
target.  The dissociation constant is largely defined by the
sequence specificity of the TF to the
promoter~\cite{Bintu2005}, and thus it is subject to evolutionary
selection~\cite{Ronen2002,Zaslaver2004,Yosef2011}.  
If  noise in the timing is a variable with
phenotypic relevance, the threshold level that triggers the target
response may have been tuned so as to, for instance, minimise the
noise in the FPT.  
Given a steady-state protein level $p_{ss}$ that
the regulator dynamics will reach asymptotically, the system could
select a threshold $\alpha$ with the smallest noise in the time
``measured'' by the target response.  The question is how the FPT
variability depends on the threshold level $\alpha$, and if an optimal
strategy for controlling timing noise actually exists.

In the case of simple activation without further regulation ("constitutive"), the noise in the protein
number is known analytically (Eq.~\ref{cvp}), and thus
Eq.~\ref{cvtzero} can be used as a first estimate of the FPT noise.
In this approximation, the relative fluctuations in the crossing time
are a function of the frequency $a$ and size $b$ of expression bursts,
and of the threshold $\alpha$,

\begin{equation}
 CV_t^2 = \frac{\alpha}{a~b} \frac{1+b(2-\alpha)}{[(1-\alpha)log(1-\alpha)]^2}.
\label{cvtconst}
 \end{equation}

 This approximate analytical expression shows a non-monotonous dependence on $\alpha$, 
 and is in very good agreement with the results of exact 
 stochastic simulations of the process (Fig.~\ref{fig4}A).  The noise
 in the FPT has a minimum at an intermediate level of protein
 expression, implying that the threshold position can actually be
 selected in order to minimise the noise in the ``time measurement''.
 Intuitively, the presence of this minimum is the result of a
 trade-off between the noise at the protein level and the steepness of
 the increase in time of the average protein level, as described by
 Eq.~\ref{cvtzero}.  Both quantities are decreasing functions of the
 threshold position in the case $\eta \ll 1$. This can be easily
 observed by taking the time derivative of the average protein
 dynamics (Eq.~\ref{sol:mean_field}) and by simply reformulating the
 protein noise in Eq~\ref{cvtzero} as

\begin{equation}
 CV_p^2(\alpha) = \frac{1+b(2 -\alpha )}{a~b~\alpha}.
 \end{equation}
 
 For short times the protein noise is high, while for long times the
 noise propagation is particularly efficient.  The optimal trade-off
 is in the intermediate region.  
 The timing noise  depends on the specifics of the gene in analysis, i.e.,  
  on the burst size $b$ and frequency $a$ (Eq.~\ref{cvtconst}). 
 However, the value of $\alpha$  at which this noise has a minimum shows
  no dependence on the transcription rate, thus
 on the burst frequency $a$ for a fixed protein lifetime, and a dependence on the burst size $b$
 that is completely negligible in the regime $b\gg 1$, which is 
 typically the case empirically as described in the Materials and Methods section.  In fact, in
 this regime the minimum position takes a constant value, 
 approximately halfway to the steady state:

\begin{equation}
 \frac{dCV_t^2(\alpha)}{d\alpha} \Big|_{b\gg1} = 0 \Rightarrow \alpha \simeq 0.55.
 \label{cvmin}
  \end{equation}
This value compares well to simulations (Figure~\ref{fig4}A).  

As expected, the protein noise level determines the absolute value of
the timing fluctuations. In fact, if the same number of proteins is
produced with burst sizes of different amplitudes both the noise at
the protein level (Eq.~\ref{cvp}) and the timing noise
(Eq.~\ref{cvtzero} and Fig.~\ref{fig4}A) increase.

Note that if the
process is simply a birth-death Poisson process, which can be the
relevant case if the regulator is a non-coding RNA, the FPT noise has
analogously a minimum value, in this case at $\alpha=0.47$.  This can
be calculated by  substituting the steady-state value $ab$ with the
corresponding steady state $k/\gamma$ of the birth (rate $k$) and
death (rate $\gamma$) process, and taking the limit $b\rightarrow 0$
in Eq.~\ref{cvtconst}.

In the specific case of a burst size much larger than the threshold level (i.e., $b\gg\tilde{p}$),
 the problem of evaluating the FPT fluctuations greatly simplifies. In fact, in this specific case, 
 the first transcription event nearly always leads to a burst of protein production that crosses the threshold. 
 Therefore, the FPT distribution is well approximated by an exponential distribution with a single parameter $k_m$ 
 (i.e., the transcription rate) as can be easily tested with stochastic simulations.

We considered so far stable proteins with a halflife longer than the cell cycle, 
which is typically the case in microorganisms~\cite{Taniguchi2010,Shahrezaei2008}.  
However, few proteins, such as stress response regulators, are actively degraded by proteolyis~\cite{Gottesman1996}, 
and protein halflife can be controlled in synthetic circuits~\cite{Cameron2014}.  
For unstable proteins, the higher degradation rate  $\gamma_p$ can be compensated by 
an increased transcription rate $k_m$ or translation rate  $k_p$ in order to reach the same 
steady-state level of expression $ab$.
In the first case, both the burst size $b=k_p/\gamma_m$ and frequency $a=k_m/\gamma_p$ remain constant, 
while in the second case the burst size increases to compensate the reduced frequency.    
Eqs.~\ref{cvtconst} shows that, for a fixed steady state $ab$, the minumum noise level (i.e., for $\alpha~\simeq0.55$)   
does not change for unstable proteins transcribed more often, while timing become more noisy if the burst 
size is increased. 
The average FPT corresponding to the minimum noise level decreases as the protein becomes less stable since 
$t\simeq - log(1-\alpha)/\gamma_p$. Therefore, if a 
signal has to be transmitted reliably on very short time scales with respect to the cell cycle, a possible cellular strategy is 
boosting the protein degradation at the cost of making more transcripts to achieve a 
shorter average crossing time without an increase in its relative fluctuations.
%

On the other hand, given a fixed
average value of the crossing time, one can ask whether the timing
variability can be reduced by the cell by ``paying the cost'' of
producing more proteins.  Eq.~\ref{cvtconst} shows that increasing the
transcription rate, and thus the burst frequency $a$, can indeed
decrease the absolute value of fluctuations. On the other hand, the
timing variability does not depend on the burst size for $b\gg1$.
Therefore, making more proteins in order to have a more precise
crossing time is a possible strategy if the expression is increased at
the transcriptional level, but does not work  if the increase occurs at the translational level
(Fig.~\ref{fig4}B).

\subsection*{Role of autoregulation in controlling timimg fluctuations}

This section
addresses how the timing noise is affected by self-regulation of the gene.
Gene autoregulation is widespread in both bacteria and eukaryotes and
has relevant consequences on the gene expression dynamics and
stochasticity~\cite{CosentinoLagomarsino2007,Alon2007}.  For example,
 negative transcriptional self-regulation
speeds up the expression rise-time after
induction~\cite{Rosenfeld2002} and can reduce the cell-to-cell 
variability  in the protein number  at steady state~\cite{Dublanche2006}, while  
positive autoregulation slows down the time of
induction~\cite{Maeda2006}, and increases stochastic fluctuations
eventually leading to expression bimodality in a specific range of
parameters~\cite{Maeda2006,Becskei2001}.

Transcriptional regulation is described by multiplying the transcription rate of the target with a
nonlinear function $F(p)$ of the level of expression of the
regulator~\cite{Alon2007,Bintu2005}. The empirical dependence
is well captured by a Hill function~\cite{Bintu2005a} of the form:
\begin{eqnarray}
\hspace*{-0.6cm}  F_{-}(p) ~&=&~ \frac{1}{[1+(p/K)^n ]} ~~~\textrm{for negative autoregulation} ,  \nonumber\\
\hspace*{-0.6cm}  F_{+}(p) ~&=&~ \frac{(p/K)^n }{[1+(p/K)^n ]} ~~~\textrm{for positive autoregulation}.
\label{hill}
\end{eqnarray}
Here, the dissociation constant $K$ specifies the regulator level at which
the production rate is half of its maximum value, while the Hill
coefficient $n$ defines the degree of cooperativity of the regulator
and thus the steepnes of the regulation curve.

\begin{figure*}
  \centering
 \includegraphics[width=1.\textwidth]{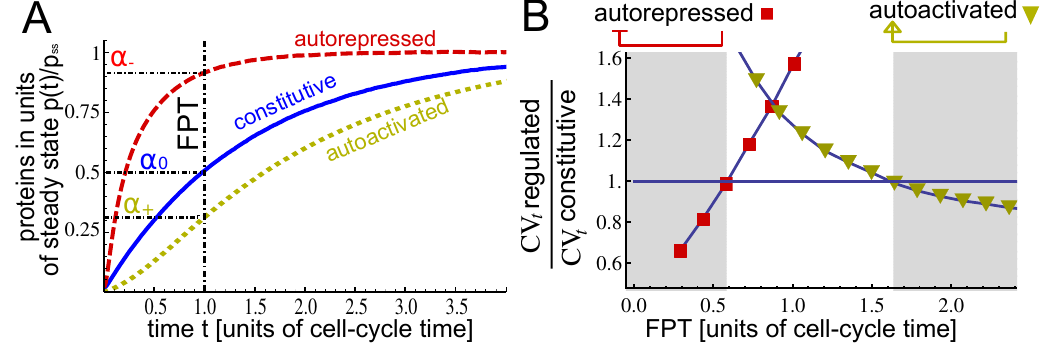}
 \caption{ \textbf{Timing noise can be reduced by different regulatory
     strategies in different time regimes.} A) Setting of the
   comparison between regulatory strategies: we imposed the same
   steady state of expression for the three different circuits.   A
   common average FPT (vertical line) is achieved by placing the target protein level
   $\alpha$ at a position that depends on the circuit type, corresponding to $\alpha_{0,+,-}$ in this example.
   Protein and mRNA lifetimes are defined as 5 min and 8 hours, 
   while the steady-state number of proteins is $a b = 2000$ as in Fig.~\ref{fig4}. 
   The average burst size has a value of $b=5$.
   The strengths of
   regulation are defined by the dissociation constants $K=1000$ and
   $K=90$ (proteins) for negative and positive autoregulation
   respectively.  The Hill exponent $n=2$ was used for regulations.
   B) With the constraints shown in A, the timing noise for
   autorepression (squares) and autoactivation (triangles) are
   compared to the timing noise of a constitutive gene  as a function of the average FPT value.  
   Given the  common final steady-state level of expression, 
   each  average  FPT  (x-axis) corresponds to a 
   different threshold level for the 
   three circuits (as illustrated in panel A) and to a different noise level.  
   When the mean    FPT is close to the protein halflife (or to the doubling time for
   stable proteins) both transcriptional autoregulations increase the
   timing noise (central region). The two shaded regions show the FPTs
   for which self-regulation works as a timing noise filter.  In
   particular, autorepression reduces timing noise for time scales
   shorter than about 0.5 doubling times, while autoactivation is
   efficient in timing noise reduction on time scales longer than
   about 1.5 doubling times.
}
  \label{fig5}
\end{figure*}

To compare in a meaningful way different regulatory strategies, it is
essential to precisely define the constraints and the criteria to put different circuits on
equal footing~\cite{Alon2007,Bosia2012}.  For example, to show that negative transcriptional
regulation speeds up the gene response to activation, its dynamics can
be compared to the one of a constitutive gene with the same
steady-state level of expression~\cite{Alon2007,Rosenfeld2002}.  
  We
adopt the same approach, comparing different
regulatory strategies while keeping fixed the final expression level.  This can be achieved in practice by choosing the
appropriate values of the regulation strengths (defined by the
dissociation constants $K$ in Eqs.~\ref{hill}) and of the basal
transcription rates $k_m$ for the three circuits.  Given this
constraint, the same average response time can be achieved by
different types of autoregulation (positive or negative) or by a
constitutive promoter by setting the crossing threshold at different
positions.  This is shown in Fig.~\ref{fig5}A: the protein expression
tends asymptotically to the same equilibrium level in the three
circuits, and the same average FPT (vertical line) is measured by placing different
thresholds ($\alpha_{0,+,-}$)  because of the different dynamics of the average protein level $p(t)$ in the three circuits.

Figure~\ref{fig5}B shows the ratio between the timing noise of the two
self-regulations and the timing noise for constitutive expression.
Depending on the average crossing time, different regulatory
strategies have different noise properties.  Around the time scale set
by the protein halflife (cell-cycle time in bacteria) adding any type of autoregulation introduces larger
timing fluctuations.  On the other hand, if the crossing time is
shorter than the cell cycle, smaller timing fluctuations can be
achieved introducing negative self-regulation, while positive
autoregulation reduces the timing noise for longer time scales.  The
scenario emerging from this result is that the type of regulation
that can buffer timing fluctuations crucially depends on the time that
the cell has to precisely ``measure'' with respect to the typical time
scale of the process (protein halflife or cell-cycle time for stable proteins).  The reduction or
amplification of noise in the protein number at the steady state seems
to be more directly associated to the sign of autoregulation,
i.e., noise reduction for repression and noise amplification for
activation~\cite{Alon2007a}. Instead, the noise properties of
autoregulation are context dependent at the level of FPT fluctuations
(Fig.~\ref{fig5}B).

This effect can be understood qualitatively by looking at
the dynamics of the protein level in Figure~\ref{fig5}A, and considering
Eq.~\ref{cvtzero} to connect noise at the protein level and timing noise. 
Now, the dynamics of  $p(t)$
is strongly dependent on the type of regulation.  
The expression of a
constitutive gene crosses an intermediate (with respect to the steady
state) protein level at a time  close to the protein halflife
since $t\simeq - log(1-\alpha)/\gamma_p $, and this is the range where
its timing noise is close to a minumum.  On the other hand, for
negative autoregulation this same timing corresponds to a protein
level much closer to the steady-state value, where the derivative
$dp(t)/dt$ is much smaller, and thus the fluctuations are strongly
amplified at the timing level.  The opposite is true for short times,
where this derivative has a larger value if the gene is negatively
autoregulated.  Analogous  observations hold for
positive autoregulation.  Clearly, at a quantitative level, the fact
that autoregulation can significantly change the noise at the protein
level plays an important role.  However, the contribution from the deterministic dynamics seems 
dominant in most cases.

As previously discussed, positive self regulation can display bistability, and thus bimodality, in the protein level for specific parameter values~\cite{Maeda2006,Becskei2001}. 
In this particular case, the protein expression level can converge to a stable no- or low-expression state or to a high-expression state depending on initial conditions. 
Assuming a threshold value between these two stable states and an initial condition of low expression,   
the timing of threshold crossing would be defined by the residence time in the low-expression state. 
Bistable circuits are typically found at the basis of cell-fate determination and phenotypic heterogeneity, for 
example in the context of bacterial persistence and competence~\cite{Norman2015}, where noise-driven stochastic switches are relatively rare. 
Therefore, a bistable circuit seems a less well suited strategy to transmit signals within timescales comparable to the cell cycle and with a reliable timing,  
as in cell-cycle regulation or circadian clocks. As a consequence, we did not explore parameter regimes for which the positive feedback shows bistability.  
In this case, other mathematical tools can be used to estimate the fluctuations in the residence times~\cite{Walczak2005,Assaf2011}.

\section{DISCUSSION}


Our two most important results are the following. First, the
fluctuations of the time necessary to reach a threshold expression level 
have a minimum
value. This optimal threshold level of expression is approximately half
the steady-state value.  Thus, it does not naively coincide with a
noise minimum in the protein number, which is instead monotonously
decreasing while approaching the steady-state level.  As a
consequence, the level of noise in protein number does not directly
translate into a level of timing fluctuations, since the
out-of-equilibrium dynamics plays a non-negligible role.  Therefore,
the cellular strategies to control noise can be significantly
different depending on what is exactly the biologically relevant
variable, e.g., protein intracellular concentration or
threshold-passing time.

Second, positive and negative transcriptional gene self-regulation can
alter the level of timing fluctuations: the former reduces timing
fluctuations at short times compared to one cell cycle (the system's
intrinsic time scale), while the latter reduces timing noise at large
times.
In other words, the role of autoregulation in controlling timing noise
is context dependent.  In the intermediate region (around one cell
cycle), the absence of any regulation is the best strategy.
Importantly, different regulatory strategies have been compared in a
mathematically controlled way~\cite{Alon2007}, thus measuring the
timing variability given a fixed average crossing time and a fixed
number of proteins produced.  The cell can in principle further reduce
the timing noise, as well as the noise at the protein level, by paying
the cost of producing more proteins. However, we showed that also for
timing noise this is effective only if the production is increased at
the transcription level.

Giving up the ambition of a full analytical solution for the FPT
problem, we provide a general approximate relation linking timing
fluctuations and noise at the protein level that leads to a
closed-form expression for the timing noise of constitutive
expression.  This expression was not employed in previous studies; we
tested it with exact stochastic simulations, and appears to be useful in
many regimes.
Importantly, this simple estimate may be applicable to cellular
first-passage time noise problems in more general contexts than the
gene expression problem considered here, including protein
modification, signal transduction, and titration, which has likely
importance for cell-cycle progression~\cite{Amodeo2016,Schmoller2015a}
The importance of this estimate is that it gives a clear intuition
about general features of timing fluctuations.  
%
%
In particular, in our case we can capture and explain the fact that the
timing noise shows a minimum value at a threshold expression level
approximately half of the steady-state value.  In the case of a
transcription factor inducing a target gene, this translates into
setting the value of the dissociation constant around half the
steady-state level of the regulator.  This general property can be
used for the design of synthetic genetic circuits such as clocks and
oscillators for which a precise timing is relevant~\cite{Tigges2009}.


A recent interesting study also produced analytical estimates of the
FPT fluctuations~\cite{Singh2014}, but only in the specific case of gene
expression in absence of protein degradation or dilution through
growth.  Although the results appear to be relevant for the case of
lysis time variation in the bacteriophage lambda, they do not easily
generalise to the classic scheme of gene expression in
Fig.~\ref{fig1}, which is more realistic for most genes.  Indeed, as
we showed, the intrinsic time scale defined by protein halflife plays
an important role in shaping timing fluctuations. 


Another study~\cite{Shahrezaei2008} proposed an approach to
estimate the FPT distribution based on numerical solutions of a
renewal equation, but did not explored systematically the sources or
controls of timing variability.  A subsequent work~\cite{Murugan2011},
focused on the role of the mRNA to protein lifetime (the parameter
$\eta=\gamma_p/\gamma_m$) in shaping timing fluctuations for
autoregulatory loops.  Using a continuous approximation and numerical
simulations, the authors showed that in these circuits timing
fluctuations can be tuned by $\eta$. 
Here, we extend the list of the key variables determining timing
fluctuations, and we show that different autoregulatory strategies can
efficiently control timing noise at different time scales. 
Importantly, we also provide an analytical explanation of our results, 
based on a simple, although approximate, approach to evaluate timing 
fluctuations and intuitively understand some of its general properties.  

While we only considered self-regulatory strategies,   
 more complex genetic circuits can play an important role in shaping timing fluctuations. 
A  recent theoretical analysis suggests for example that incoherent feedforward loops 
can be an efficient topology to dampen timing fluctuations~\cite{Ghusinga2016}.


An important technical extension of the work presented here would be
the inclusion of extrinsic fluctuations, i.e., fluctuations in the
model parameters due to the variability of global factors affecting
gene expression such as polymerases or ribosome
concentrations~\cite{Swain2002}.  Extrinsic noise can be relevant
especially for highly expressed genes~\cite{Taniguchi2010}, and its
effects on timing fluctuations is completely unknown.  Unfortunately,
the sources of extrinsic noise are still not fully understood, and
analytical calculations easily become unfeasible in presence of
parameter fluctuations.  Therefore, this extension is out of the scope
of the present work and will require an extensive exploration of the
possible sources and functional forms of extrinsic noise and of its
propagation at the timing level.
Moreover,  more complex models explicitly accounting for cell-cycle progression, DNA replication and cell division    
may be necessary to fully capture the details of expression timing fluctuations, 
especially if the time scales in analysis are 
short with respect to the cell cycle~\cite{Bierbaum2015,Soltani2016}. 

%


Importantly, fluctuations in the time necessary for a regulator to reach a
threshold  expression level  may not be the only source of the overall timing noise.  In fact, 
 a regulator reaching a critical expression level also needs to  be ``sensed'' by the downstream  processes. 
 For example, TFs have to find and bind target promoters in order to control their expression. 
 This ``reading process''  takes some time and introduces additional timing variability.   
 An order-of-magnitude estimate of reaction times points to the presence of a separation of time scales between the 
 relatively slow change in a protein level due to transcription and translation  
 and the faster kinetics of target search and binding/unbinding to promoters~\cite{Alon2007,Swift2016}. 
More precisely,  single-molecule  experiments addressing the kinetics of TF search~\cite{Li2009}, have shown that 
 a single TF molecule can find its binding site in about 4 minutes~\cite{Hammar2012},  and this search time is expected to reduce 
 proportionally to the number of TFs~\cite{Slutsky2004}. Hence, for TFs that are present in hundreds or thousands of copies, 
 we expect these times to be negligible compared to changes
in protein levels, which happen on a time scale of the order of the
cell cycle.
  These observations support a prominent role of the regulator expression dynamics 
  with respect to the reading process of its targets in establishing timing fluctuations, 
  unless the TF copy number is very low and the threshold crossing time is relatively fast.  
  This general consideration finds some experimental evidence in the context of
yeast meiosis: the variability in the expression dynamics of
a meiotic master TF was shown to be the dominant source 
of variability in the onset time of downstream targets (and
thus on the resulting phenotipic variability)~\cite{Nachman2007}.
However, a generalization of our modelling framework to include  downstream processes should be required in cases in which 
 the out-of-equilibrium kinetics of TF binding  plays a non negligible role in the observed 
 expression dynamics~\cite{Hammar2014}.


While the biologically relevant circuits are in general more complex
than those considered here, we can speculate on the implications of
our results in a wider context.
It is interesting to notice that the promoter of the dnaA gene is
repressed by its own protein DnaA, forming a negative
self-loop~\cite{Grant2011}.  DnaA is responsible for the initiation of
DNA replication in \textit{E.~coli} by promoting the unwinding of the
double-strand DNA when its level reaches a certain
threshold~\cite{Skarstad2013}.  The initiation timing have to be
precisely regulated to couple cell growth and division with DNA
replication~\cite{Adiciptaningrum2015}, and this timing is clearly
shorter than the cell-cycle time. Our analysis suggests that indeed
negative self-regulation can help controlling the timing fluctuations
on such a time scale.

Additionally, the importance of expression timing could be relevant
for genes beyond cell-cycle and circadian clock regulators.  In fact,
several endogenous genes are expressed in a precise temporal order.
This is the case for example for genes involved in flagellar
biosynthesis~\cite{Kalir2001}, or genes coding for enzymes in the
aminoacid biosynthesis systems of
\textit{E. coli}~\cite{Zaslaver2004}.  This temporal order has been
proposed as the result of an optimization process due to a trade-off
between speed and cost of production~\cite{Alon2007,Zaslaver2004}.
Often the genetic network implementing this temporal order is composed
by a single TF that triggers the response of a set of target genes at
different threshold levels~\cite{Alon2007}.  If the delay between the
expression of these genes has to be tightly tuned, timing fluctuations
in reaching the different thresholds could be
detrimental. Interestingly, many of these master TFs are
autoregulated.  For example the lrhA gene, a key regulator controlling
the transcription of flagellar, motility and chemotaxis genes, shows
positive autoregulation~\cite{Lehnen2002}.  On the other hand, some of
the pathways in aminoacid synthesis are controlled by a single
regulator with a negative self-loop~\cite{Zaslaver2004}.  It is
tempting to speculate that this can be also due to regulation of
timing noise and not only of its average value. In fact, the response
of metabolic genes is known to be generally fast (order of
minutes)~\cite{Zaslaver2004}, while the complete formation of a
functioning flagella is intuitively a more time consuming process.
Ideally, experiments directly checking the timing variability for
different values of the threshold, for example looking at the response
in fluorescence of target promoters with different binding affinities,
or measurements of the timing noise in presence or absence of
autoregulation would be the perfect tests of our theoretical work.

\section{Funding} 
The work was partially supported by the Compagnia San Paolo grant GeneRNet.

\section{ACKNOWLEDGEMENTS}

We thank Matthew AA Grant for useful discussions.

\subsubsection{Conflict of interest statement.} None declared.

\bibliographystyle{nar.bst}
\bibliography{FPT.bib}

\end{document}